\documentclass[a4paper,11pt]{article}
\usepackage{pos}
\usepackage{siunitx}
\usepackage{xfrac}

\usepackage{subcaption}
\usepackage{tikz}
\usetikzlibrary{babel, calc, arrows, angles, quotes, patterns, patterns.meta, shapes.geometric, intersections, decorations.pathmorphing, decorations.markings}

\usepackage{tabularray}
\usepackage{booktabs}
\UseTblrLibrary{booktabs}
\UseTblrLibrary{siunitx}
\usepackage{makecell}

\usepackage[version=4, math-greek=default, text-greek=default]{mhchem}

\usepackage[size=tiny]{todonotes}
\presetkeys{todonotes}{fancyline, color=blue!30}{} 

\usepackage{bibspacing}
\title{Gradient-descent-based reconstruction for muon tomography based on automatic differentiation in PyTorch}
\ShortTitle{Gradient-descent-based reconstruction for muon tomography}

\author*[a]{Jean-Marco Alameddine}
\author[a]{Felix Sattler}
\author[a]{Maurice Stephan}
\author[a]{Sarah Barnes}

\affiliation[a]{German Aerospace Center, Institute for the Protection of Maritime Infrastructures,\\
  Fischkai 1, Bremerhaven, Germany}

\emailAdd{jean-marco.alameddine@dlr.de}

\renewcommand{\logo}{\relax}

\abstract{
Muon scattering tomography is a well-established, non-invasive imaging technique using cosmic-ray muons.
Simple algorithms, such as PoCA (Point of Closest Approach), are often utilized to reconstruct the volume of interest from the observed muon tracks.
However, it is preferable to apply more advanced reconstruction algorithms to efficiently use the sparse muon statistics that are available.
One approach is to formulate the reconstruction task as a likelihood-based problem, where the material properties of the reconstruction volume are treated as an optimization parameter.

In this contribution, we present a reconstruction method based on directly maximizing the underlying likelihood using automatic differentiation within the PyTorch framework.
We will introduce the general idea of this approach, and evaluate its advantages over conventional reconstruction methods.
Furthermore, first reconstruction results for different scenarios will be presented, and the potential that this approach inherently provides will be discussed.
}

\FullConference{Fifth MODE Workshop on Differentiable Programming for Experiment Design (MODE2025)\\
8-13 June 2025\\
Kolymbari, Crete, Greece\\}

\begin{document}
\maketitle

\section{Introduction}

Muon tomography is a powerful imaging technique which exploits the properties of naturally occurring cosmic-ray muons.
The Earth is constantly hit by high-energy particles, notably protons and heavier atomic nuclei, originating from the cosmos.
As these particles interact with the Earth's atmosphere, cascades of secondary particles are initiated.
From these cascades, especially muons are capable of reaching the Earth's surface, leading to a nearly constant flux of one muon per \si{\square\centi\meter} per \si{\minute}~\cite{Grupen:437495}.
Due to their large mass, which is more than \num{200} times higher that the mass of an electron, as well as the absence of strong interactions, muons possess a tremendous penetrating power, making them highly valuable for tomographic tasks.

As muons pass through a medium, they interact with atomic nuclei through Coulomb scattering, leading to a deflection from their normally straight path.
Since the magnitude of these deflections is correlated to the properties of the traversed material, mainly the mass density and the atomic number, it is possible to measure the muons before and after passing an unknown object, and obtain material information from the observed directional changes.
This method is called muon scattering tomography.
Compared to alternative imaging techniques, such as X-ray scanning systems, muon scattering tomography offers several advantages:
Since muons occur naturally, no active sources are needed, which means that no additional radiation exposure for humans and animals is necessary, reducing the bureaucratic effort of dealing with radiation protection and eliminating any possible health hazards.
Furthermore, the penetrating power of muons allows for the examination of heavily shielded materials, which poses a problem for traditional methods.
Both points make muon scattering tomography especially interesting for an application in the security domain, for example for contraband detection at border crossings, airport, or harbors~\cite{instruments7010013}.

Since muon detectors are well-understood and commercially available~\cite{instruments7010013}, the most important remaining challenge in muon scattering tomography is the most efficient usage of the limited available muon statistics to be able to reconstruct materials with adequate accuracy in an acceptable time.
Simple geometric reconstruction approaches, such as the Point of Closest Approach (PoCA) algorithm~\cite{10.1063/1.1606536} or the Angle Statistics Reconstruction (ASR) algorithm~\cite{Stapleton_2014}, are inherently limited in their capabilities due to their underlying simplifications.
On the other hand, more advanced statistical methods~\cite{schultz} are computationally expensive and often numerically unstable, requiring careful tuning of these methods.

In this contribution, a likelihood-based reconstruction method for muon scattering tomography is presented.
The main idea of this algorithm is the direct minimization of the negative log-likelihood of the problem, using the automatic differentiation framework provided by the Python machine learning library PyTorch~\cite{paszke2019pytorchimperativestylehighperformance}.
This workflow allows for the utilization of modern optimizing techniques provided by PyTorch, as well as an easy path to extend the algorithm in the future.
After introducing and formulating the reconstruction task as a likelihood problem in Section~\ref{sec:methodology}, the implementation of the algorithm is detailed in Section~\ref{sec:results}, together with the presentation of first reconstruction results based on simulation data.
Lastly, future prospects of the presented approach are outlined in Section~\ref{sec:outlook}.

\section{Methodology}
\label{sec:methodology}

\subsection{Reconstruction Task in Muon Tomography}
\label{sec:reco_task}

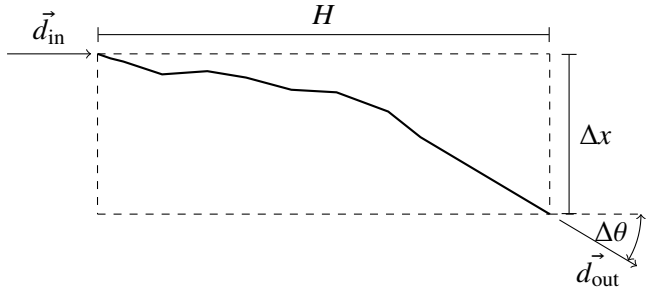
\begin{figure}[h]
	\centering
    \begin{tikzpicture}[scale=0.85, every node/.style={scale=1.0}]
        \centering
\coordinate (origin) at (0,0);
\coordinate (p1) at (8,0);
\coordinate (p2) at (0, -3);

\draw [->] (-1.4, 0) -- node[above] {$\vec{d_\text{in}}$} (-0.1, 0);

\coordinate (scatter_end) at (7, -2.5);
\coordinate (penultimate_point) at (5.5, -1.6);
\draw [thick] (origin) -- (0.2, -0.07) -- (0.4, -0.12) -- (1, -0.32) -- (1.7, -0.27) -- (2.3, -0.37) -- (3, -0.56) -- (3.7, -0.6) -- (4.5, -0.9) -- (5, -1.3) -- (penultimate_point) -- (scatter_end);

\draw [->] ($(scatter_end) + 0.1*(scatter_end) - 0.1*(penultimate_point)$) -- node[below] {$\vec{d_\text{out}}$} ($(scatter_end) + 0.9*(scatter_end) - 0.9*(penultimate_point)$);

\coordinate (anchor1) at ($(scatter_end) + (1, 0)$);
\coordinate (anchor2) at ($(scatter_end) + (scatter_end) - (penultimate_point)$);
		\draw pic["$\Delta \theta$", draw=black,<->,angle eccentricity=0.7,angle radius=1.2cm] {angle=anchor2--scatter_end--anchor1};

\draw [|-|] ($(origin) + (0, 0.3)$) -- node[above] {$H$} ($(scatter_end |- 42,0) + (0, 0.3)$);
\draw [|-|] ($(scatter_end |- 42,0 ) + (0.3, 0)$) -- node[right] {$\Delta x$} ($(scatter_end) + (0.3, 0)$);

\draw [dashed] (origin) -- (scatter_end |- 42,0);
\draw [dashed] (0, 42 |- scatter_end) -- (scatter_end);
\draw [dashed] (origin) -- (0, 42 |- scatter_end);
\draw [dashed] (scatter_end |- 42,0) -- (scatter_end);
\draw [dashed] (scatter_end) -- ($(scatter_end) + (1.4, 0)$);
    \end{tikzpicture}
    \caption{Visualization of the multiple scattering variables $\Delta \theta$ and $\Delta x$. Adapted from \cite{alameddine2024}.}
    \label{fig:scattering_sketch}
\end{figure}

When a muon traverses a medium, its direction is continuously altered due to multiple scattering.
For a muon covering a distance $H$, multiple scattering can be defined via the variable $\Delta \theta$, describing the angular change between the initial direction $\vec{d_\text{in}}$ and the final direction $\vec{d_\text{out}}$, and the variable $\Delta x$, describing the displacement of the muon.
Both variables are visualized in Figure~\ref{fig:scattering_sketch}.
Accurately describing the distributions of $\Delta \theta$ and $\Delta x$ is complex.
Therefore, a commonly used simplification is to describe them as a joint Gaussian distribution~\cite{Workman:2022ynf}, which provides a good approximation of small scattering angles.
In this case, the likelihood of a deflection $\Delta \theta$ after a distance $H$ is given as~\cite{Workman:2022ynf, schultz}
\begin{equation}
     \label{eq:gaussian}
     f(\Delta\theta) \, \mathrm{d}(\Delta\theta) = \frac{1}{\sqrt{2 \pi} \sigma_\theta} \exp{\left( - \frac{\Delta \theta^2}{2 \sigma_\theta^2} \right)} \, \mathrm{d}(\Delta\theta),
\end{equation}
with
\begin{align}
    \label{eq:variance}
    \sigma_{\Delta \theta}^2 &= \lambda H (p_0 / p)^2, & \lambda \equiv \left( 15 / p_0 \right)^2 1 / L_\text{rad},
\end{align}
where $p$ is the momentum of the muon, $p_0$ is a reference momentum, $\lambda$ is the scattering density, and $L_\text{rad}$ is the material-dependent radiation length.
The variance of the Gaussian distribution describing $\Delta x$ and its correlation with $\Delta \theta$ are given as~\cite{Workman:2022ynf}
\begin{align}
    \sigma_{\Delta x}^2 &= \frac{H²}{3} \sigma_\theta^2, & \rho_{\Delta \theta \Delta x} &= \frac{\sqrt{3}}{2}.
\end{align}
This means that within this approximation, the traversed medium is only characterized by $\lambda$.
Obtaining the distribution of $\lambda$ within a medium, given the measured scattering of muons, is the reconstruction task at hand.
For this purpose, the scanned medium is often discretized into voxels, where $\lambda$ is assumed to be constant in each voxel.
The obtained map of $\lambda$ can either be used directly for material identification, or fed to subsequent analysis steps such as anomaly detection.

One of the simplest geometric approaches to perform this reconstruction task is the PoCA algorithm, which works under the assumption that the entire scattering along the muon path occurred at a single point~\cite{10.1063/1.1606536}.
This simplification restricts the capability of the approach to reconstruct complex scenes, and it leads to an inefficient use of the limited available statistics since every muon contributes scattering information only to a single voxel.
The ASR algorithm~\cite{Stapleton_2014} tries to improve the latter by assuming that the scattering occurred in an extended region around the point of closest approach.
While adding additional information to the reconstruction, this approach leads to a typical elongation of the reconstruction results along the direction of the ingoing muons.
As an alternative to these geometric methods, statistical methods can be used:
Based on the probability distributions of $\Delta x$ and $\Delta \theta$ in~\eqref{eq:gaussian}, a likelihood $\mathcal{L}$ is formulated.
This likelihood, which will be introduced in Section~\ref{sec:likelihood}, describes how well the voxel map of $\lambda$ is capable of describing the observed scattering data.
In~\cite{schultz_phd}, minimizing this likelihood directly by optimizing $\lambda$ has been successfully attempted, however, the approach was found to be unsuitable for real-time application due to high computational demands.
Subsequently, an approach to optimize the likelihood via an iterative expectation maximization algorithm has been developed first in~\cite{schultz}.

\subsection{Likelihood Formulation}
\label{sec:likelihood}

The measurable observables in muon scattering tomography are the positions and directions of the muon when entering and when exiting a VOI (volume of interest).
From this, the data vector $\mathbf{D}_i = (\Delta \theta_i, \Delta x_i)^T$ is defined for each muon $i$, where the definitions of $\Delta \theta_i$ and $\Delta x_i$ are visualized in Figure \ref{fig:scattering_sketch}.
In analogy to the Gaussian multiple scattering distribution in~\eqref{eq:gaussian}, the likelihood of $\lambda$ in a voxelized VOI for a single muon event is given as~\cite{schultz}
\begin{equation}
    \label{eq:single_likelihood}
    \mathcal{L}(\mathbf{D}_i | \lambda) = \frac{1}{2 \pi \sqrt{\left| \mathbf{\Sigma}_i \right|}} \exp \left( - \frac{1}{2} \mathbf{D}_i^T \mathbf{\Sigma}_i^{-1} \mathbf{D}_i \right),
\end{equation}
with the covariance matrix
\begin{equation}
    \label{eq:covariance_matrix}
    \mathbf{\Sigma}_i = (p_0 / p_i)^2 \sum\limits_{j} \left( \lambda_j \mathbf{W}_{ij} \right),
\end{equation}
where the index $j$ runs over all voxels that are hit by the muon path.
It is important to note that the muon trajectory within the VOI is unknown, which means that assumptions about the real muon path based on the measured ingoing and outgoing positions need to be made.
Two possible assumptions are the \emph{straight line path}, where the known ingoing and outgoing positions are connected by a straight line, and the \emph{PoCA path}, where the muon path is drawn from the ingoing muon position to the point of closest approach to the outgoing muon position.
While $p_0$ is a fixed reference value, as defined in \eqref{eq:variance}, $p_i$ is the absolute muon momentum per event.
For $p_i$, usually no or only rough estimations are available since a precise estimation on a per-event basis is currently not feasible.
In many cases, $p_i$ is therefore set to a value that does not necessarily represent reality.
The weight matrix $\mathbf{W}_{ij}$ contains the geometric information about the assumed voxel path, and is defined by~\cite{schultz}
\begin{equation}
    \label{eq:weight_matrix}
    \mathbf{W}_{ij} = \begin{bmatrix} L_{ij} & L_{ij}^2 / 2 + L_{ij} T_{ij} \\ L_{ij}^2 / 2 + L_{ij} T_{ij} & L_{ij}^3 / 3 + L_{ij}^2 T_{ij} + L_{ij} T_{ij}^2 \end{bmatrix},
\end{equation}
where $L_{ij}$ denotes the distance muon $i$ covers in voxel $j$, and $T_{ij}$ the distance of muon $i$ from the end of voxel $j$ until the end of the VOI.
From the likelihood $\mathcal{L}(\mathbf{D}_i | \lambda)$ for a single muon, the negative log likelihood for $M$ muons is obtained via
\begin{align}
    \label{eq:likelihood}
        -\log\left( \mathcal{L} \right) &= - \log \left(  \prod_{i}^{M}  \mathcal{L} (\mathbf{D}_{i,x} | \lambda) \cdot  \mathcal{L} (\mathbf{D}_{i,y} | \lambda)  \right) \nonumber \\
        &= - \sum_{i}^{M} \Biggl[ \log \left( \frac{1}{2 \pi \sqrt{\left| \mathbf{\Sigma}_i \right|}} \exp \left( - \frac{1}{2} \mathbf{D}_{i,x}^T \mathbf{\Sigma}_i^{-1} \mathbf{D}_{i,x} \right) \right) +
          \log \left( \frac{1}{2 \pi \sqrt{\left| \Sigma_i \right|}} \exp \left( - \frac{1}{2} \mathbf{D}_{i,y}^T \mathbf{\Sigma}_i^{-1} \mathbf{D}_{i,y} \right) \right)  \Biggr] \nonumber \\
            &= \sum_{i}^{M} \left( \log \left( |\mathbf{\Sigma}_i| \right) + \frac{1}{2} \mathbf{D}_{i,x}^T \mathbf{\Sigma}_i^{-1} \mathbf{D}_{i,x} + \frac{1}{2} \mathbf{D}_{i,y}^T \mathbf{\Sigma}_i^{-1} \mathbf{D}_{i,y} \right) + \text{const.},
\end{align}
where $\mathbf{D}_{i,x}$, $\mathbf{D}_{i,y}$ describe the data vector projected onto the $xz$-plane, respectively the $yz$-plane.
\section{Implementation of the Method and Evaluation on Simulation Data}
\label{sec:results}

Since the negative log likelihood in~\eqref{eq:likelihood} quantifies how likely it is that the voxel map $\lambda$ describes the observed muon data, minimizing $-\log\left( \mathcal{L} \right)$ with respect to $\lambda$ can be used as the reconstruction approach.
For that, the calculation of $-\log\left( \mathcal{L} \right)$ is implemented in Python within the PyTorch framework~\cite{paszke2019pytorchimperativestylehighperformance}.
This allows for the usage of the \texttt{torch.autograd} functionality, which offers automatic differentiation:
After calculating a result (here: $-\log\left( \mathcal{L} \right)$), \texttt{torch.autograd} automatically collects the gradients from all numerical calculations with respect to a given set of parameters (here: $\lambda$), and combines them using the chain rule.
The calculated gradient is used in combination with an optimizer (here using the \texttt{adam} algorithm~\cite{adam}) over several epochs to find an optimal value for $\lambda$.

\begin{figure*}
    \centering
    \begin{subfigure}[t]{0.485\textwidth}
        \centering
        \includegraphics[trim={0 0 0 0cm},clip,width=\textwidth]{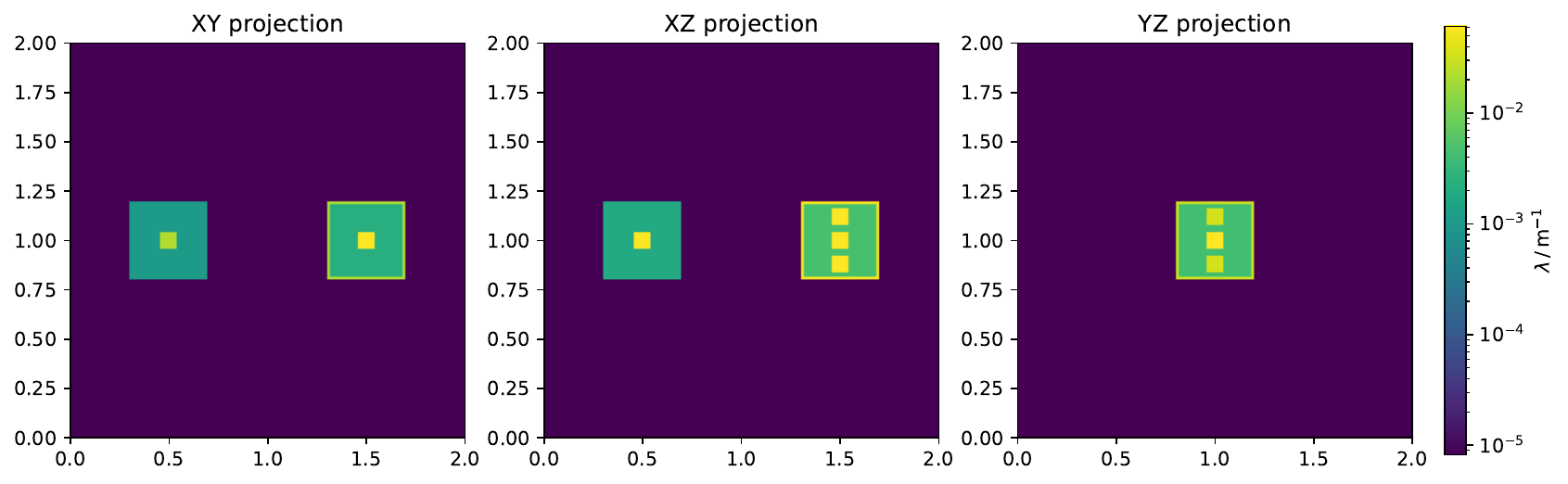}
        \caption[]%
        {{\small Ground truth of the scene. The left object is a water cube, the right object is a hollow iron box. The small cubes within are made out of tungsten.}}
        \label{fig:gt_asrtest}
    \end{subfigure}
    \hfill
    \begin{subfigure}[t]{0.485\textwidth}
        \centering
        \includegraphics[trim={0 0 0 1.5cm},clip,width=\textwidth]{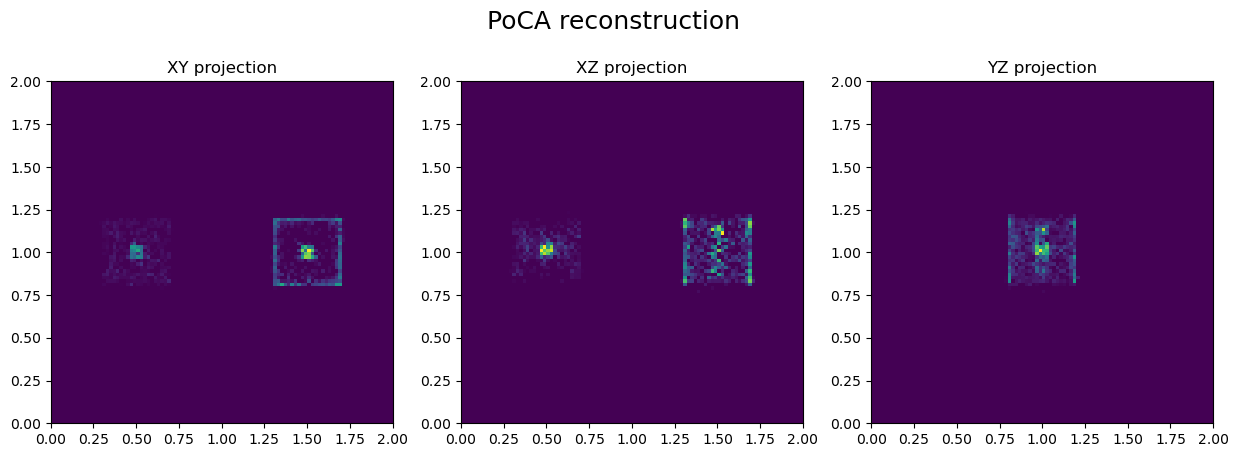}
        \caption[]%
        {{\small Reconstruction result of the scene using the PoCA algorithm~\cite{10.1063/1.1606536}.}}
        \label{fig:poca_asrtest}
    \end{subfigure}
    \vspace{-0.3cm}\vskip\baselineskip
    \begin{subfigure}[t]{0.485\textwidth}
        \centering
        \includegraphics[trim={0 0 0 1.5cm},clip,width=\textwidth]{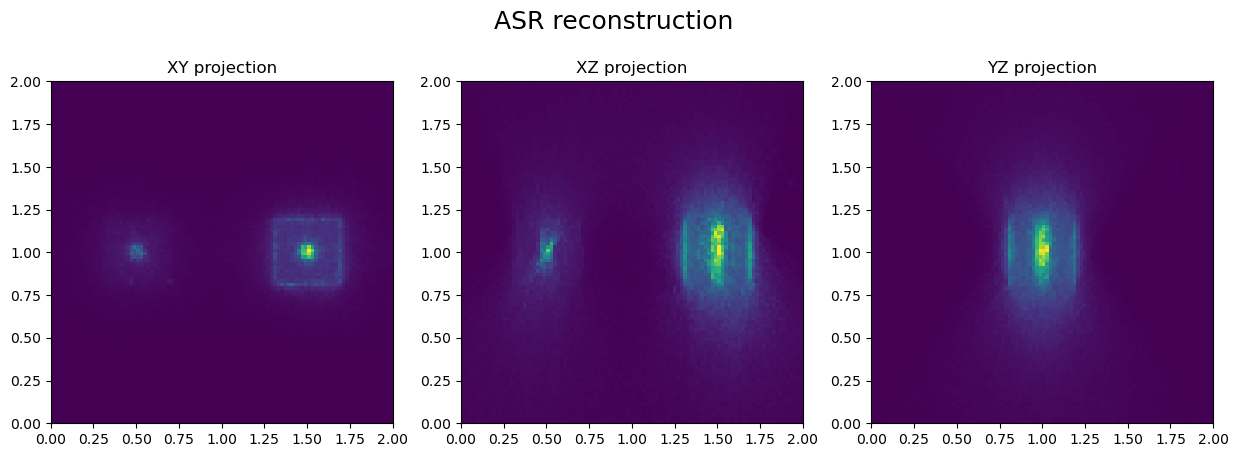}
        \caption[]%
        {{\small Reconstruction result of the scene using the ASR algorithm~\cite{Stapleton_2014}.}}
        \label{fig:asr_asrtest}
    \end{subfigure}
    \hfill
    \begin{subfigure}[t]{0.485\textwidth}
        \centering
        \includegraphics[trim={0 0 0 1.5cm},clip,width=\textwidth]{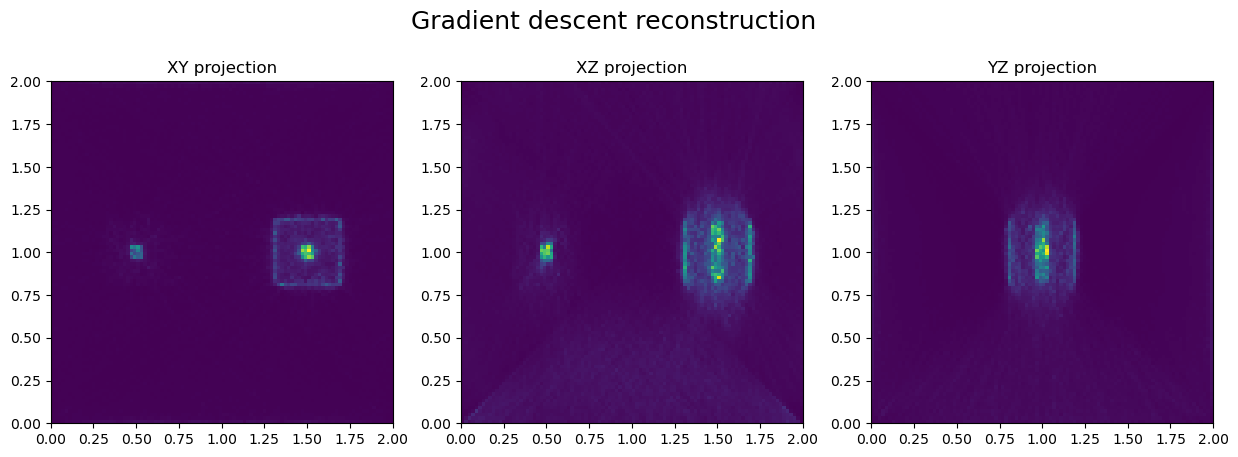}
        \caption[]%
        {{\small Reconstruction result of the scene using the gradient descent method presented in this work.}}
        \label{fig:grad_asrtest}
    \end{subfigure}
    \caption[ The average and standard deviation of critical parameters ]
    {Evaluation of different reconstuction algorithms based on simulation data for the first scene.}
    \label{fig:reco_asrtest}
\end{figure*}

\begin{figure*}
    \centering
    \begin{subfigure}[t]{0.485\textwidth}
        \centering
        \includegraphics[trim={0 0 0 0},clip,width=\textwidth]{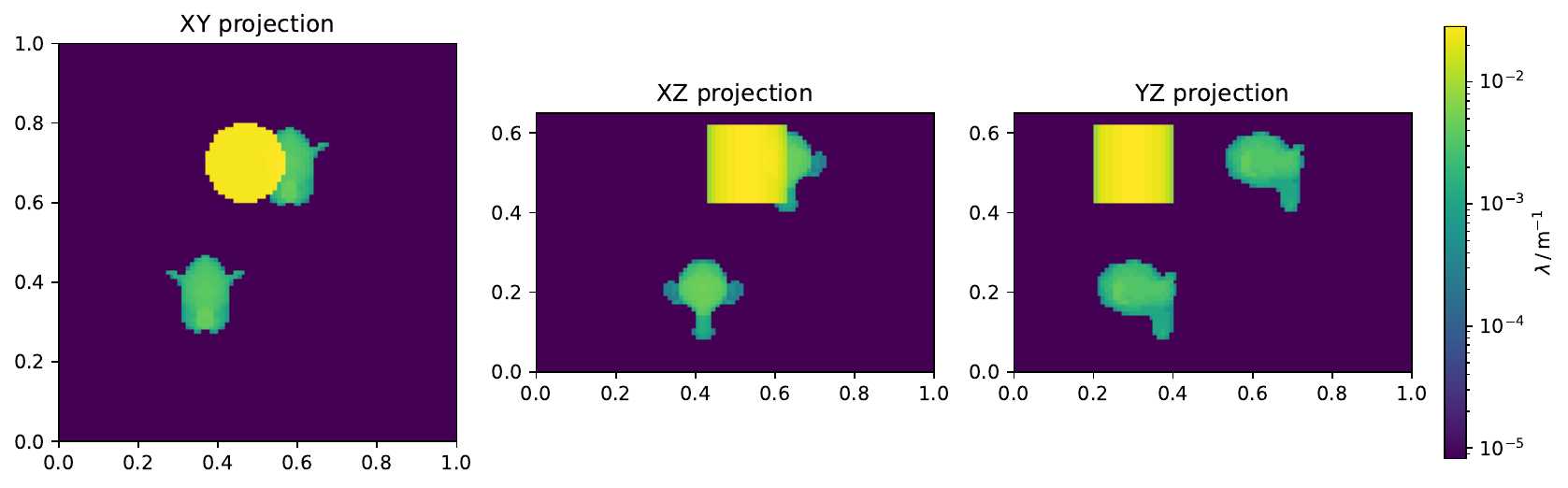}
        \caption[]%
        {{\small Ground truth of the scene. The cylinder is made of iron, while the two sculptures are made of $\ce{CaCO3}$.}}
        \label{fig:gt_monkey}
    \end{subfigure}
    \hfill
    \begin{subfigure}[t]{0.485\textwidth}
        \centering
        \includegraphics[trim={0 0 0 1.5cm},clip,width=\textwidth]{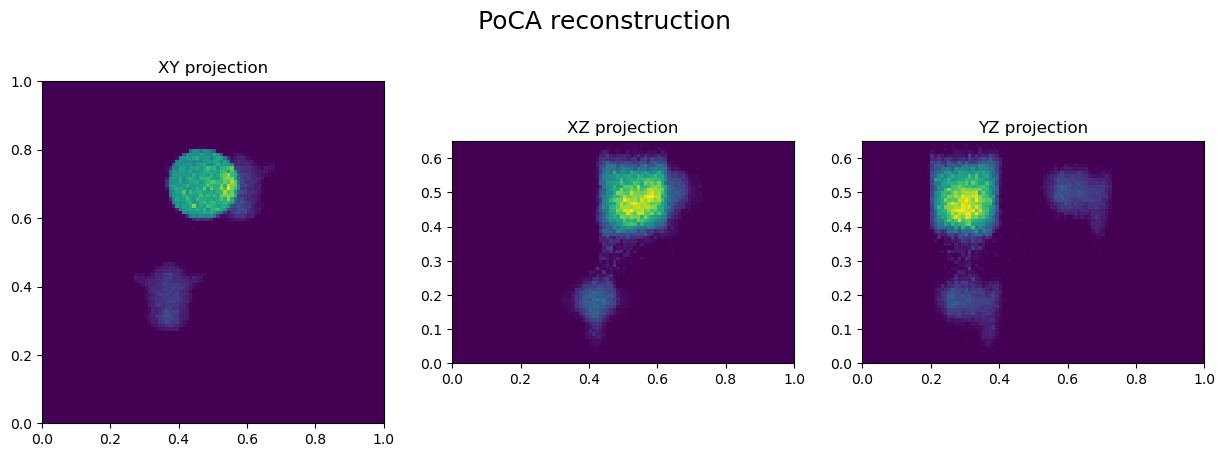}
        \caption[]%
        {{\small Reconstruction result of the scene using the PoCA algorithm~\cite{10.1063/1.1606536}.}}
        \label{fig:poca_monkey}
    \end{subfigure}
    \vspace{-0.3cm}\vskip\baselineskip
    \begin{subfigure}[t]{0.485\textwidth}
        \centering
        \includegraphics[trim={0 0 0 1.5cm},clip,width=\textwidth]{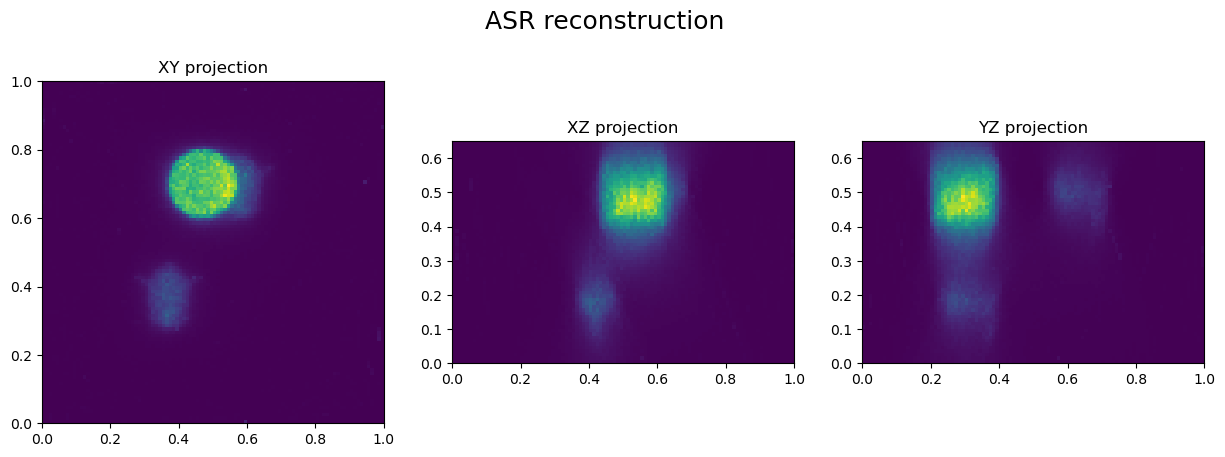}
        \caption[]%
        {{\small Reconstruction result of the scene using the ASR algorithm~\cite{Stapleton_2014}.}}
        \label{fig:asr_monkey}
    \end{subfigure}
    \hfill
    \begin{subfigure}[t]{0.485\textwidth}
        \centering
        \includegraphics[trim={0 0 0 1.5cm},clip,width=\textwidth]{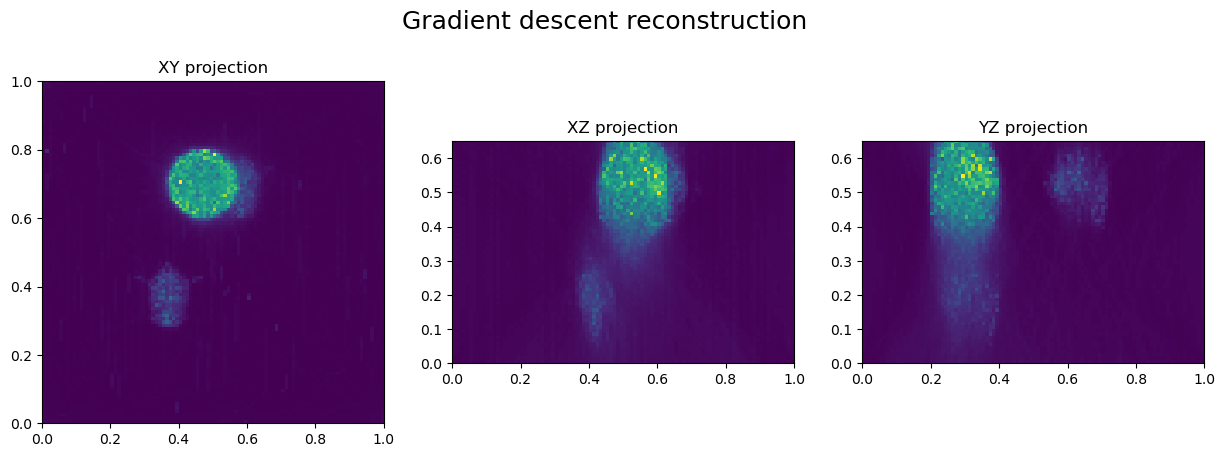}
        \caption[]%
        {{\small Reconstruction result of the scene using the gradient descent method presented in this work.}}
        \label{fig:grad_monkey}
    \end{subfigure}
    \caption[ The average and standard deviation of critical parameters ]
    {Evaluation of different reconstuction algorithms based on simulation data for the second scene.}
    \label{fig:reco_monkey}
    \vspace{-0.5cm}
\end{figure*}

To validate this method, two muon tomography simulation datasets are created using realistic cosmic-ray muon distributions generated with the CRY library~\cite{4437209}, muon propagation performed with the \textsc{Geant4} toolkit~\cite{AGOSTINELLI2003250}, and 3D scenes generated with the B2G4 framework for usage within \textsc{Geant4}~\cite{b2g4}.
The first scene, see Figure~\ref{fig:gt_asrtest}, features a water cube and an iron box, each containing smaller tungsten cubes.
The second scene, see Figure~\ref{fig:gt_monkey}, consists of three randomly distributed solid objects.
These scenes are chosen since they represent scenarios where imaging is challenging, involving objects that block or enclose each other.
For each scene, \num{1.5e7} muons are injected on a $\SI{10}{\meter} \times \SI{10}{\meter}$ plane, corresponding to a measurement time of \SI{15}{\minute}.
The detectors have a dimension of $\SI{2}{\meter} \times \SI{2}{\meter}$ with a detector distance of \SI{1}{\meter} for the first scene, respectively $\SI{1}{\meter} \times \SI{1}{\meter}$ and \SI{0.65}{\meter} for the second scene, where a perfect detector resolution is assumed.

The reconstruction results using the PoCA algorithm~\cite{10.1063/1.1606536}, the ASR algorithm~\cite{Stapleton_2014}, and the gradient descent reconstruction presented in this work are shown in Figure~\ref{fig:reco_asrtest} and Figure~\ref{fig:reco_monkey}.
For the gradient descent approach, the algorithm is stopped after \num{25} epochs, since no significant improvement in the reconstruction result is visible afterward.
As an initialization, all voxels are initialized with $\lambda = \lambda_\text{air}$, which means that no prior information is introduced.
Furthermore, the algorithm is performed in a mini-batch gradient descent approach:
This means that for each epoch, the dataset is split randomly into $M_\text{batch}$ batches (here: $M_\text{batch} = 15$, corresponding to \num{e6} events per batch), and the optimization step is performed separately for each batch.
The path assumption made for the voxel tracing is the \emph{PoCA path} approximation, as explained in Section~\ref{sec:likelihood}.
The momentum is set to $p = \SI{750}{\mega\electronvolt}$ in~\eqref{eq:covariance_matrix} for all muons, assuming that no momentum estimation is available, with a reference momentum of $p_0 = \SI{3000}{\mega\electronvolt}$.
These numerical values are chosen via a simple hyperparameter search since they provide the best convergence behavior.

Visually, the results show that the $xy$ projection is well reconstructed with all algorithms in both scenes.
For the $xz$ and $yz$ projections, the general shapes are clearly visible, although no algorithm is able to clearly distinguish between the individual blocks in the first scene.
In addition, the ASR algorithm contains a visible smearing along the $z$ direction, which is characteristic for this approach, while the PoCA algorithm is unable to fully reconstruct the small boxes within the first scene.
A quantitative analysis of the similarity between the reconstruction results and the ground truths is given in Table~\ref{tab:numerical_results}.
The applied metrics are the Peak Signal-to-Noise Ratio (PSNR), the Structural Similarity Index Measure (SSIM)~\cite{ssim}, and the Learned Perceptual Image Patch Similarity (LPIPS)~\cite{lpips}.
Note that for the PSNR and SSIM, higher values correspond to a higher similarity, while lower values correspond to a higher similarity for the LPIPS\@.
For the first scene, the quantitative results are very similar, with a different best-performing algorithm for each metric.
For the second scene, the gradient descent algorithm performs the best, although again the margins are very small, especially for the PSNR\@.
This result shows that the gradient descent algorithm is already capable of performing at least as good as existing standard algorithms without extensive optimization, and without the usage of its further potentials which will be outlined in Section~\ref{sec:outlook}.
\begin{table}
    \caption{Numerical evaluation of the reconstruction results using different metics. The best result for each scene and each reconstruction is highlighted in bold.}
    \centering
    \begin{tabular}{lcccccc}
        \toprule
        & \multicolumn{3}{c}{Scene 1} & \multicolumn{3}{c}{Scene 2} \\
        \cmidrule(lr){2-4} \cmidrule(lr){5-7}
        & \makecell{PSNR} & \makecell{SSIM} & \makecell{LPIPS} & \makecell{PSNR} & \makecell{SSIM} & \makecell{LPIPS} \\
        \midrule
        PoCA & \num{22.324} & \textbf{\num{0.870}} & \textbf{\num{0.477}} & \num{19.666} & \num{0.737} & \num{0.378} \\
        ASR  & \num{22.321} & \num{0.829} & \textbf{\num{0.477}} & \num{19.699} & \num{0.685} & \num{0.395} \\
        Gradient descent & \textbf{\num{22.348}} & \num{0.863} & \num{0.480} & \textbf{\num{19.960}} & \textbf{\num{0.838}} & \textbf{\num{0.326}} \\
        \bottomrule
    \end{tabular}
    \label{tab:numerical_results}
\end{table}
\begin{table}
    \caption{Computational benchmarking of the algorithms for the first scene. Note that for the gradient descent, convergence and runtime of the algorithm depend on the exact problem and the chosen hyperparameters.}
    \centering
    \begin{tabular}{lcc}
    \toprule
    & Memory consumption & Execution time \\
    \midrule
    PoCA & $\approx$\SI{400}{\mega\byte} & $\approx$\SI{1.4}{\second} \\
    ASR & $\approx$\SI{2300}{\mega\byte} & $\approx$\SI{14.5}{\second} \\
    Gradient descent & $\approx$\SI{3400}{\mega\byte} & $\approx$\SI{21.7}{\second} \\
    \bottomrule
    \end{tabular}
    \label{tab:performance}
\end{table}

A benchmarking of the memory consumption and the execution times of the algorithms is presented in Table~\ref{tab:performance}.
Both the memory consumption and the execution time are the smallest for the PoCA algorithm due to its simplicity.
In the use case presented here, the benchmark results for the ASR algorithm and the gradient descent algorithm are comparable.
Note that the memory consumption of the gradient descent algorithm can be steered via the batch size $M_\text{batch}$, since only the relevant data from each batch needs to be stored at a time.
This makes the gradient descent method scalable in terms of memory consumption.
For the scenes at hand, batch sizes between $M_\text{batch}=1$ and $M_\text{batch}=500$ all produced similar reconstruction results, validating that the approach is also viable for smaller batches where the statistical variance for each batch is larger.

\section{Further Prospects and Conclusions}
\label{sec:outlook}

In this work, a likelihood-based gradient descent reconstruction method for muon scattering tomography has been presented.
After formally introducing the likelihood, the algorithm has been implemented in PyTorch, making use of the \texttt{torch.autograd} automatic differentiation functionality to perform the optimization.
Using two created simulation datasets, it has been shown that the reconstruction results are already qualitatively and quantitatively able to compete and even surpass results obtained with traditional reconstructions methods such as the PoCA and ASR algorithms.
The required computational resources are comparable to those required by the ASR method, with scalability possible due to the mini-batch gradient descent approach.

The greatest potential of the presented method lie within the improvements to the algorithm that are intrinsically possible due to its likelihood formulation.
This allows for the introduction of penalty terms $p(\lambda)$, with $-\log\left( \mathcal{L} \right) \rightarrow -\log\left( \mathcal{L} \right) + p(\lambda) $.
With that, prior information can be introduced to the problem, for example knowledge about expected materials or geometric shapes.
As introduced in Section~\ref{sec:reco_task}, the multiple scattering distribution is described as Gaussian.
While this approximation is valid for small scattering angles, it is unable to correctly incorporate the large scattering angles caused by the tails of the real scattering distribution.
This issue can be encountered by formulating the likelihood in~\eqref{eq:single_likelihood} as a Gaussian scale mixture model~\cite{5204665},
\begin{align*}
    \mathcal{L}_\text{GSM} = \sum\limits_{k}^{} w_k g(\mathbf{\Sigma}_{i,k}) \quad \text{with} \quad g(\Sigma_{i,k}) \propto \exp \left( - \frac{1}{2} \mathbf{D}_i^T \frac{1}{s_k} \mathbf{\Sigma}_{i,0}^{-1} \mathbf{D}_i \right),
\end{align*}
where the parameters $w_k$ and $s_k$ can either be set to appropriate values, or treated as free parameters during optimization.



\bibliographystyle{JHEP}
\bibliography{references}


\end{document}